\documentclass[aps,prl,twocolumn,groupedaddress]{revtex4}

\usepackage{graphicx}

\begin{document}

\title{Magnetic-Field-Induced Mott Transition in a Quasi-Two-Dimensional Organic Conductor} 

\author{F. Kagawa,$^{1}$ T. Itou,$^{2}$ K. Miyagawa,$^{1, 2}$ and K. Kanoda$^{1, 2}$}

\affiliation{
$^{1}$Department of Applied Physics, University of Tokyo, Bunkyo-ku, Tokyo 113-8656, Japan
\\
$^{2}$CREST, Japan Science and Technology Corporation, Kawaguchi 332-0012, Japan
\\}

\date{\today}

\begin{abstract}
	We investigated the effect of magnetic field on the highly correlated metal near the Mott 
transition in the quasi-two-dimensional layered organic conductor, 
$\kappa$-(BEDT-TTF)$_{2}$Cu[N(CN)$_{2}$]Cl, by the resistance measurements under control of 
temperature, pressure, and magnetic field. 
	It was demonstrated that the marginal metallic phase near the Mott 
transition is susceptible to the field-induced localization transition of the first order, 
as was predicted theoretically. 
	The thermodynamic consideration of the present results gives a conceptual pressure-field phase 
diagram of the Mott transition at low temperatures.

\end{abstract}



\maketitle

	Through extensive experimental and theoretical studies, it has been well recognized that 
almost localized Fermions (ALF) is the source of various interesting phenomena such as 
high-$T_{\rm C}$ superconductivity (SC) and colossal magnetoresistance. 
	However, comprehension of the nature of ALF is still in progress, and many problems are 
left unsolved. 
	One of them is the effect of magnetic field on ALF. 
	In the theoretical approach to this problem, the Hubbard model with half-filled band has been 
often employed and attacked using Gutzwiller approximation earlier \cite{Ref1} and dynamical
mean-field theory recently \cite{Ref2,Ref3}. 
	The two approaches share an important prediction that ALF undergoes first-order localization 
transition by magnetic field, although different in detail. 
	One example of the experimental realization in this context is liquid $^{3}$He, 
which is often viewed as the strongly correlated liquid near the Mott localization. 
	However, the field-induced localization transition was not observed in liquid $^{3}$He up to 
200 T \cite{Ref4}. 
	It is argued that the discrepancy between the experimental and theoretical results is
attributable to inadequacy of the lattice description of liquid $^{3}$He through the Hubbard model 
with half filling \cite{Ref3}.

\begin{figure}
\includegraphics[width=7.5cm,height=11.5cm]{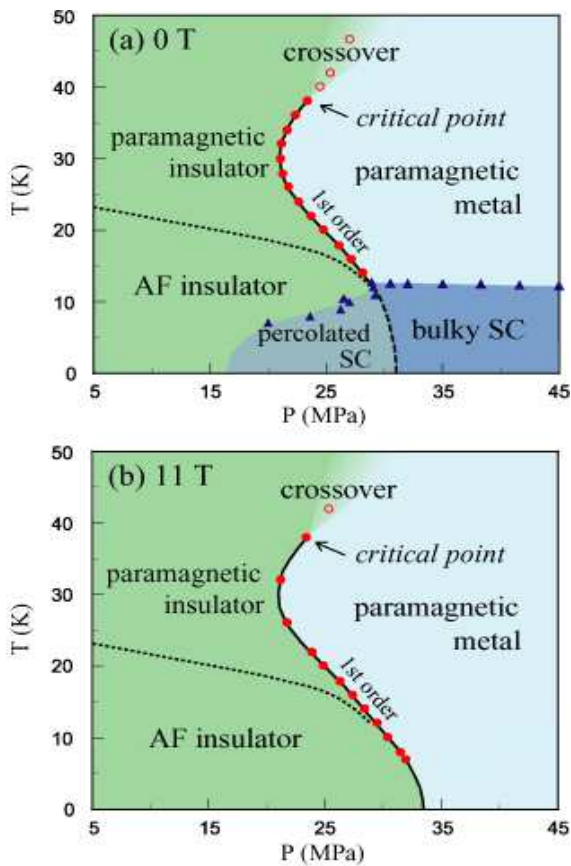}
\caption{\label{Fig1} Pressure-temperature phase diagrams of $\kappa$-Cl under a zero
field (a) taken from Ref. [9] and under a field of 11 T normal to the conducting layer (b). 
	Closed and open circles represent points giving the first-order Mott transition 
and crossover, respectively. 
The AF transition lines (dotted line) and the AFI-SC boundary 
(broken line) are taken after Lefebvre \textit{et al}. [6].}

\end{figure}

	The highly correlated electronic system with half-filled band provides another experimental 
stage suitable for the study of this issue. 
Recently, it has been established that the quasi-two-dimensional (quasi-2D) layered organic conductor, 
$\kappa$-(BEDT-TTF)$_{2}$Cu[N(CN)$_{2}$]Cl (denoted by $\kappa$-Cl hereafter), is a prototype of 
the bandwidth-controlled Mott transition system with a half-filled single band 
\cite{Ref5,Ref6,Ref7,Ref8,Ref9}. 
	The pressure-temperature ($P$-$T$) phase diagram of $\kappa$-Cl under a zero field is shown in 
Fig. \ref{Fig1}(a) which is taken from our previous work \cite{Ref9}. 
	Under pressure, $\kappa$-Cl shows the first-order Mott transition from the paramagnetic insulator 
(PI) to the paramagnetic metal (PM) with a critical endpoint at $\sim$ 38 K. 
	Thus the present Mott transition is viewed as a genuine one, which is driven only by 
electron-electron correlation and doesn't accompany symmetry breaking, as posutulated by Mott 
\cite{Ref10}.

	The purpose of the present work is to investigate the effect of magnetic field on the highly 
correlated metal near the Mott transition, namely a sort of ALF, in the quasi-2D system, $\kappa$-Cl. 
	We performed the resistivity measurements for $\kappa$-Cl under control of temperature, 
pressure, and magnetic field. 
	The results evidenced that the marginal metallic phase undergoes 
the field-induced localization transition of the first order, as is predicted theoretically.

	First, in order to establish the $P$-$T$ phase diagram under a high magnetic field above the 
upper critical field, $H_{\rm C2}$, we performed the in-plane resistance measurements at 11 T 
(normal to the conducting layer) under isothermal $pressure$ $sweep$ by use of He gas. 
	The $\kappa$-Cl crystal used here is the identical sample used in our previous work \cite{Ref9}.
	The $P$-$T$ phase diagram under 11 T obtained in the present experiment is shown in 
Fig. \ref{Fig1}(b), where closed and open circles represent the first-order Mott transition characterized
by jump in resistance and crossover points defined by a pressure giving 
maximum in pressure-derivative of logarithmic resistance, $|\frac{1}{R}\frac{\partial R}{\partial P}|$, respectively. 
	The AF order lines (dotted line) in Figs. \ref{Fig1} (a) and (b) are depicted identically, 
because NMR measurements on the fully deuterated salt, $\kappa$-d$_{8}$-(BEDT-TTF)$_{2}$Cu[N(CN)$_{2}$]Br, 
which is just on the border of the Mott transition, indicated that the N\'eel temperature is field-insensitive
up to 12 T at least \cite{Ref11}.
%

	Above $\sim$ 13 K ($\sim T_{\rm C}$) the first-order lines under the different
 fields are nearly the same (see Fig. \ref{Fig1}); the pressure-induced transition above
 $\sim$ 13 K under 11 T is considered to be from PI to PM as in the zero field case. 
	However, there is small but meaningful difference in a 
range of $\sim$ 13 K $<$ $T$ $<$ $\sim$ 24 K: the pressures of the Mott transition 
under 11 T are slightly higher than in the zero field case by $\sim$ 0.2 MPa at 14.1 K 
and $\sim$ 0.1 MPa at 20.1 K, 
although the differences are not appreciable in the scale of Fig. \ref{Fig1}. 
	Below 13 K, the affect of magnetic field is prominent because a magnetic field suppresses 
the bulky SC phase in higher pressure side and the minor SC domains \cite{Ref12} in the predominant 
antiferromagnetic insulator (AFI) in lower pressure side \cite{Ref9}. 
	Thus, in the phase diagrams, percolated SC phase and bulky SC phase under a zero field are 
replaced by AFI phase and PM phase under 11 T, respectively. 
	The pressure-induced transition below $\sim$ 13 K under 11 T is considered to be from AFI to 
PM, while it is from AFI to SC under a zero field \cite{Ref6}. 
	The shift between the AFI-SC boundary [broken line in Fig. \ref{Fig1}(a)] and the AFI-PM 
boundary [solid line below $\sim$ 13 K in Fig. \ref{Fig1}(b)] is clear even in the scale of Fig. 
\ref{Fig1}.

	The feature that the first-order line shifts to the higher pressure side by 
a magnetic field implies that the marginal PM or SC undergoes the field-induced localization 
transition to PI or AFI. 
	In order to demonstrate the transition, we traced the resistance behavior under $field$ $sweep$ 
(normal to the conducting layer) with temperature and pressure kept constant. 
	The field sweep was performed at an extremely slow rate ($\sim$ 0.04 T/min) after $\kappa$-Cl 
was adjusted at a specific point, ($P^{\ast}$, $T^{\ast}$).
%

\begin{figure}
\includegraphics[width=6cm,height=11.5cm]{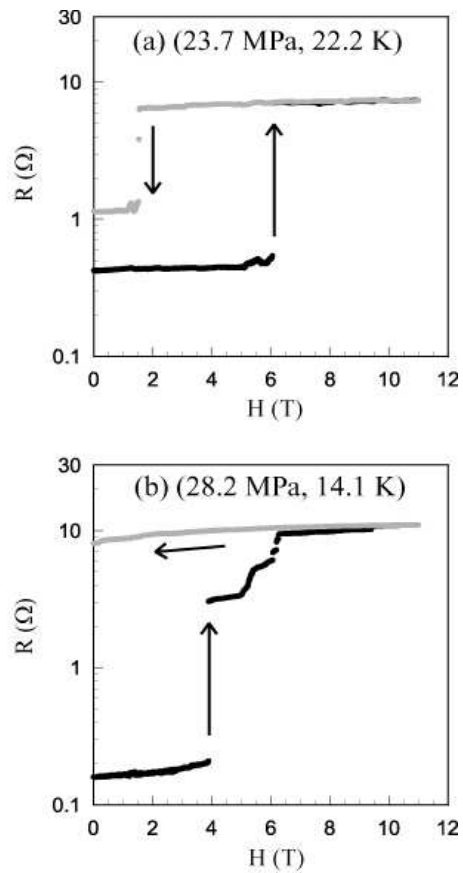}
\caption{\label{Fig2}Resistance curves under ascending (black) and descending 
(gray) fields normal 
to the conducting layer with temperature and pressure kept constant.}

\end{figure}

\begin{figure}
\includegraphics[width=8cm,height=7.5cm]{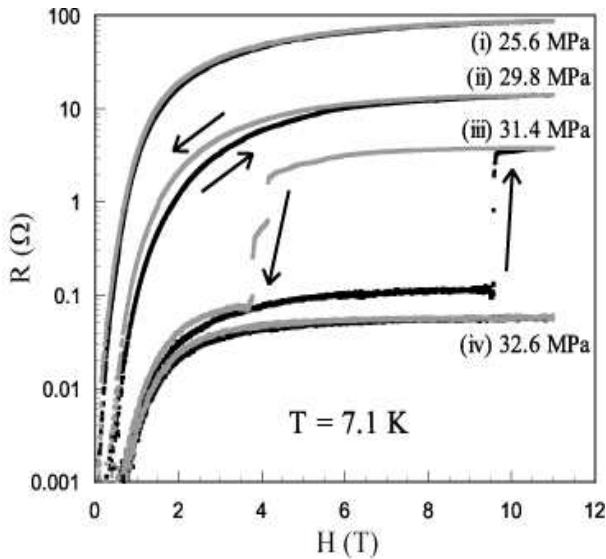}
\caption{\label{Fig3}Magnetic field dependence of resistance under various pressures at 7.1 K. 
	The field is applied normal to the conducting layer
The black and gray points are for ascending and descending fields, 
respectively.}

\end{figure}

	In a range of $\sim$ 13 K $<$ $T$ $<$ $\sim$ 24 K, we set ($P^{\ast}$, 
	$T^{\ast}$) quite near the Mott transition boundary in PM region. 
	The typical field dependence of the resistance is represented by the data  
shown in Fig. \ref{Fig2}, which demonstrates the field-induced resistive transition of the 
first order clearly by the huge jump of nearly one order of magnitude and large hysteresis. 
	The magnitude of the field of the resistance jump was very sensitive to the system 
position, ($P^{\ast}$, $T^{\ast}$), as expected. 
	The $P$-$T$ phase diagrams indicate that the field-induced 
transition at 22.2 K [Fig. \ref{Fig2}(a)] is the localization transition from PM to PI. 
        We emphasize that at 22.2 K a magnetic field induces a transition in the charge degree of 
freedom holding the spin symmetry, which is often broken concomitantly with field-induced transport 
transition. 
	Thus the transition shown in Fig. \ref{Fig2}(a) is the field-induced  
Mott transition without symmetry breaking.
        Above $\sim$ 24 K, the field-induced transition was not observed,
although we chose $P^{\ast}$ as close to the Mott transition boundary as possible 
with $T^{\ast}$ fixed at 24.9 K for example \cite{Ref13}. 
	This tendency is consistent with the feature of the phase diagrams that the first-order lines 
have no appreciable field-dependence above $\sim$ 24 K.
	Below 13 K, we performed the measurements under four pressure points at 7.1 K: (i) 
(25.6 MPa, 7.1 K), (ii) (29.8 MPa, 7.1 K), (iii) (31.4 MPa, 7.1 K), 
and (iv) (32.6 MPa, 7.1 K). 
	The field dependence of the resistance is shown together in Fig. \ref{Fig3}. 
	On the basis of the $P$-$T$ phase diagram, 
the zero-field and 11-T states at each pressure are identified as
summarized in Table I (refer to this table in the following).

	The SC phase away from the AFI-SC boundary [see Fig. \ref{Fig3}(iv)] shows a 
typical second-order transition to PM phase.
	In addition to this transition, the SC phase near the boundary
 [see Fig. \ref{Fig3}(iii)] shows a first-order transition with a huge hysteresis. 
 	In the ascending field process, the resistance jump is observed at 
 $\sim$ 9.5 T, which is far beyond $H_{\rm C2}$ at 7.1 K. Thus the first-order 
transition is the localization transition from PM to AFI. 
	At a lower pressure of 29.8 MPa [see Fig. \ref{Fig3}(ii)] across the AFI-SC boundary, 
small hysteresis without the resistance jump is observed. 
	This implies distributed first-order transition either from non-bulky SC to AFI or through 
PM (see below), which likely comes from inhomogeneous internal pressure in the sample. 
At a further lower pressure of 25.6 MPa [see Fig. \ref{Fig3}(i)], the  magnetoresistace gets to 
occur at lower magnetic fields without resistance jump nor hysteresis. 
	As is to be seen in Fig. \ref{Fig4} below, marginal SC remaining slightly in the host AFI by distributed 
internal pressure is considered to undergo SC-to-AFI transition at low magnetic fields. 
	Invisibility of the expected hysteresis is possibly due to quite small fraction of the minor 
SC domains.

	In what follows, we discuss the observed field-induced transitions in Fig. \ref{Fig2} 
(above $T_{\rm C}$) and Fig. \ref{Fig3} (below $T_{\rm C}$) from the thermodynamic point of view. 
	For clarity, we consider the pressure-field ($P$-$H$) phase diagram at a constant temperature. 
	In the same manner of the Clausius-Clapeyron relation, the slope of the first-order line, 
$dH/dP$, in the $P$-$H$ phase diagram is given as
\begin{equation}
	\frac{dH}{dP} =
	\frac{V_{\rm PI(AFI)}-V_{\rm PM(SC)}}{M_{\rm PI(AFI)}-M_{\rm PM(SC)}} = \frac{\Delta V}{\Delta M},
\end{equation}
%
where $V$ and $M$ indicate the system volume and magnetization respectively, and $\Delta V$ and 
$\Delta M$ are their jumps across the transition from PM (or SC) to PI (or AFI). 
	The $\Delta V$ is always positive because the insulating phase 
is located in the lower pressure side, and is assumed to be field-insensitive up to 11 T. 
	We first discuss the case below $T_{\rm C}$. Since AFI in $\kappa$-Cl shows weak ferromagnetism 
\cite{Ref14} and SC shows large diamagnetism of the type-II SC under magnetic fields below 
$H_{\rm C2}$, $\Delta M (= M_{\rm AFI} - M_{\rm SC}$) is positive and its magnitude should decrease 
with magnetic field increased up to $H_{\rm C2}$. 
	From Eq. (1), that leads to positive and increasing slope with a magnetic field. 
	Above $H_{\rm C2}$, $\Delta M$ is still positive because AFI 
phase keeps weakly ferromagnetic, and therefore the slope, $dH/dP$, remains positive. 
	Thus one can draw a conceptual $P$-$H$ phase diagram at $T$ $\sim$ 7 K 
(well below $T_{\rm C} \sim$ 13 K) as is shown in Fig. \ref{Fig4}, where the solid line represents 
the first-order transition and the percolated SC phase is omitted for simplicity. 
	Figure \ref{Fig4} indicates that magnetic field can induce either SC-AFI transition or SC-PM-AFI transition if $P^{*}$ is chosen appropriate and that the Mott transition shifts to the higher pressure
side with magnetic field. 
	These features are consistent with the experimental results shown in Fig. \ref{Fig3} 
and with the shift of the first-order lines below 13 K in Fig. \ref{Fig1}.

\begin{table}
\caption{\label{Table1}Zero-field and high-field states at each 
pressure in Fig. 3.}
\begin{ruledtabular}
\begin{tabular}{ccc}
$(P^{\ast},T^{\ast})$ & under 0 T & under 11 T \\ \hline
(i) (25.6 MPa, 7.1 K) & percolated SC & AFI \\
(ii) (29.8 MPa, 7.1 K) & percolated SC & AFI \\
(iii) (31.4 MPa, 7.1 K) & bulky SC & AFI \\
(iv) (32.6 MPa, 7.1 K) & bulky SC & PM 
\end{tabular}
\end{ruledtabular}
\end{table}

	On the other hand, it is difficult to predict the $P$-$H$ phase diagram at higher temperatures 
above 13 K because the sign of $\Delta M (= M_{\rm PI} -  M_{\rm PM}$) in the paramagnetic region 
is not known. 
	However, Fig. \ref{Fig2}(a), which demonstrates the field-induced PM-to-PI transition, 
indicates that the slope, $dH/dP$, should be positive at least under fields of several Tesla, 
i.e., that PI should show larger magnetization than PM according to Eq. (1). 
	The tendency that it is difficult to observe the field-induced transition above $\sim$ 24 K 
means that the slope, $dH/dP$, is much steeper, i.e., that $\Delta M$ above $\sim$ 24 K 
is smaller than below that. 
	Then, why does $\Delta M$ become appreciable below $\sim$ 24 K? 
	Two possibilities are conceivable. 
	One, which is in an approach from high temperatures, is that the growth 
of $\Delta M$ reflects the establishment of Fermi liquid nature from the bad metal
 with temperature decreased in PM phase. As is discussed in Ref. [15], the highly correlated 
metal is possibly viewed as the bad metal near the critical point but as Fermi liquid at 
sufficiently low temperatures. 
	According to the $P$-$T$ phase diagram obtained by Limelette {\it et al}. \cite{Ref8}, 
the bad metal regime along the Mott transition boundary goes into Fermi liquid regime below 
$\sim$ 25 K. 
	That supports this scenario. 
	The other possibility, which is in an approach from low temperatures, is that the large
positive $\Delta M$ below $T_{\rm C}$ owing to the SC diamagnetization survives barely
above $T_{\rm C}$ by the SC fluctuations. 
	In relation to this, we note that the pseudogap is present up to
 two times $T_{\rm C}$ in the marginal PM as was 
evidenced by $\kappa$-d$_{8}$-(BEDT-TTF)$_{2}$Cu[N(CN)$_{2}$]Br \cite{Ref16}.

\begin{figure}
\includegraphics[width=6cm,height=5cm]{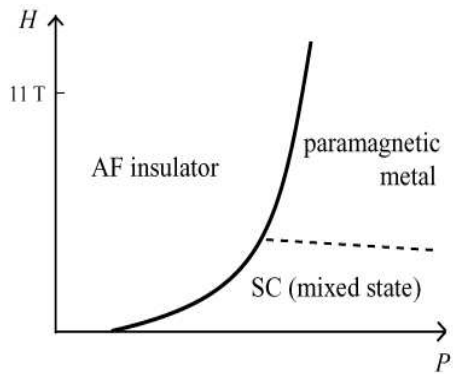}
\caption{\label{Fig4}Conceptual pressure-field phase diagram at $T$ $\sim$ 7 K, which is well below 
$T_{\rm C} \sim$ 13 K. 
	The solid line represents the first-order transition.}
\end{figure}

	To conclude, we investigated the effect of magnetic field on the 
highly correlated metal near the Mott transition in the quasi-two-dimensional organic conductor, 
$\kappa$-(BEDT-TTF)$_{2}$Cu[N(CN)$_{2}$]Cl, by resistance measurements under control of temperature, 
pressure, and magnetic field. 
	The $P$-$T$ phase diagram under 11 T normal to the conducting layer is established and found to 
have appreciable difference from the zero field case. 
	It is demonstrated that, by application of a magnetic field, (i) the marginal PM undergoes the 
Mott transition of the first order and (ii) the marginal SC undergoes 
successive transitions: the second-order SC-PM and first-order PM-AFI ones.  
	From the thermodynamic consideration incorporating the present results,
we constructed the conceptual 
%
	$P$-$H$ phase diagram of the Mott transition below $T_{\rm C}$ in the present system. 
	The present results evidenced the field-induced localization transition 
predicted from the Hubbard model, and clarified its temperature profile, which is informative
for future theoretical study.
        On the other hand, the magnetization curve in the marginal metallic phase is 
an interesting future issue, because the theories predict an upward magnetization 
curve ($\frac{\partial ^2 M}{\partial H^2} > 0$) before the field-induced transition \cite{Ref1,Ref3}.

\begin{acknowledgments}
	The authors thank S. Onoda and N. Nagaosa for fruitful discussion.
\end{acknowledgments}


\begin{thebibliography}{99}
\bibitem{Ref12} The minor SC domains near the first-order line is attributable to 
the distributed internal pressure of the crystal.
\bibitem{Ref13} After finishing the field sweep measurements at (22.4 MPa, 24.9 K), 
we depressurized the system only by 0.04 MPa, and confirmed that the resistance jump was induced.
\end{thebibliography}
\end{document}